\documentclass[aps,preprint,nofootinbib,preprintnumbers,superscriptaddress]{revtex4}

\usepackage[dvips]{color}
\usepackage[normalem]{ulem}
\usepackage{amsmath}
\usepackage{enumerate}
\usepackage{amsfonts}
\usepackage{epsfig}
\usepackage{yfonts}

\usepackage{graphicx}
\usepackage{bm}

\def\eg{{\it e.g. }}
\def\ie{{\it i.e.}}

\def\({\left(}
\def\){\right)}
\def\[{\left[}
\def\]{\right]}
\def\<{\langle}
\def\>{\rangle}

\newcommand\half{{\ensuremath{\frac{1}{2}}}}
\newcommand\p{\ensuremath{\partial}}

\newcommand\field[1]{{\ensuremath{\mathbb{{#1}}}}}

\newcommand\vev[1]{{\ensuremath{\left\langle{#1}\right\rangle}}}

\newcommand{\RR}{\field{R}}

\newcommand{\ZZ}{\field{Z}}

\newcommand{\be}{\begin{equation}}
\newcommand{\ee}{\end{equation}}
\newcommand{\bea}{\begin{eqnarray}}
\newcommand{\eea}{\end{eqnarray}}
\newcommand{\bwt}{\begin{widetext}}
\newcommand{\ewt}{\end{widetext}}

\newcommand{\bi}{\begin{itemize}}
\newcommand{\ei}{\end{itemize}}
\newcommand{\ben}{\begin{enumerate}}
\newcommand{\een}{\end{enumerate}}
\newcommand{\bca}{\begin{cases}}
\newcommand{\eca}{\end{cases}}
\newcommand{\bln}{\begin{align}}
\newcommand{\eln}{\end{align}}
\newcommand{\bst}{\begin{split}}
\newcommand{\est}{\end{split}}

\newcommand\al{{\alpha}}
\newcommand\ep{\epsilon}
\newcommand\sig{\sigma}
\newcommand\Sig{\Sigma}

\newcommand\Lam{\Lambda}
\newcommand\om{\omega}

\newcommand\ga{{\ensuremath{{\gamma}}}}
\newcommand\Ga{{\ensuremath{{\Gamma}}}}

\newcommand\De{{\ensuremath{{\Delta}}}}

\newcommand\ov{\over}
\newcommand\ha{{\half}}

\def\le{\left}
\def\ri{\right}

\newcommand\sD{{\ensuremath{{\mathcal D}}}}

\newcommand\sG{{\ensuremath{{\mathcal G}}}}

\newcommand\sN{{\ensuremath{{\mathcal N}}}}
\newcommand\sO{{\ensuremath{{\mathcal O}}}}
\newcommand\sR{{\ensuremath{{\mathcal R}}}}

\renewcommand{\Im}{\textrm{Im}\,}

\newcommand{\vk}{{\vec k}}

\newcommand\psinorm{\boldsymbol{\psi}}

\def\vertexZ{\Lambda}

\begin{document}

%

\title{From black holes to strange metals}

\preprint{MIT-CTP/4105}

\author{Thomas Faulkner}
\affiliation{KITP, Santa Barbara, CA 93106}

\author{Nabil Iqbal}
\affiliation{Center for Theoretical Physics,
Massachusetts
Institute of Technology,
Cambridge, MA 02139 }
%

\author{Hong Liu}
\affiliation{Center for Theoretical Physics,
Massachusetts
Institute of Technology,
Cambridge, MA 02139 }

\author{John McGreevy}
\affiliation{Center for Theoretical Physics,
Massachusetts
Institute of Technology,
Cambridge, MA 02139 }

\author{David Vegh}
\affiliation{Simons Center for Geometry and Physics, Stony Brook University, Stony Brook, NY 11794-3636}

\begin{abstract}

Since the mid-eighties there has been an accumulation of metallic materials whose thermodynamic and transport properties differ significantly from those predicted by Fermi liquid theory. Examples of these so-called non-Fermi liquids include the strange metal phase of high transition temperature cuprates, and heavy fermion systems near a quantum phase transition.
We report on a class of non-Fermi liquids discovered using
gauge/gravity duality.
The low energy behavior of these non-Fermi liquids is shown to be governed by a nontrivial infrared (IR) fixed point which exhibits nonanalytic scaling behavior only in the temporal direction.
 Within this class we find examples whose single-particle spectral function and transport behavior
resemble those of strange metals. In particular, the contribution from the Fermi surface
to the conductivity is inversely proportional to the temperature.
 In our treatment these properties can be understood as being controlled by the scaling dimension of the fermion operator in the emergent IR fixed point.

\end{abstract}


\maketitle

\section{Introduction}

During the last ten years, developments in string theory have revealed surprising and
profound connections between gravity and many-body systems,
resulting in the emergence of a new paradigm for strongly coupled many-body systems.
The anti de-Sitter/Conformal Field Theory (AdS/CFT) correspondence~\cite{AdS/CFT} relates a gravity theory in a weakly curved $(d+1)$-dimensional anti-de Sitter (AdS$_{d+1}$) spacetime to a strongly-coupled $d$-dimensional quantum field theory living on its boundary. This maps questions about complicated many-body phenomena at strong coupling to solvable single- or few-body classical problems in a curved geometry. Black holes in this geometry now play a surprising and universal role in characterizing the strongly coupled dynamics of the boundary theory at finite temperature and density, a development anticipated by the discovery of Hawking and Bekenstein in the 1970s that black holes are intrinsically thermodynamic objects. For example, important dynamical insight into the thermodynamics~\cite{Gubser:1996de} and transport behavior~\cite{Kovtun:2004de}
of strongly correlated systems has been obtained from simple geometric aspects of black hole spacetimes.

\begin{figure}[h]
\begin{center}
\includegraphics[scale=0.25]{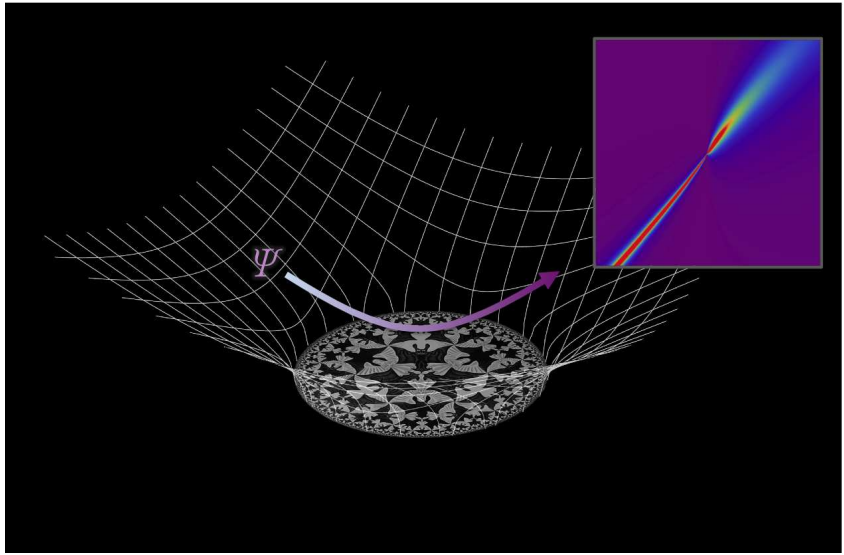}
\end{center}
\vskip -0.5cm
\caption{An illustration of the holographic approach to understanding non-Fermi liquids. The physics of a strongly-coupled many-body theory is exactly dual to that of a gravitational theory in a curved spacetime with one extra dimension. A finite density of charge in the many-body problem corresponds in the gravity theory to a black hole with an extra emergent conformal symmetry near its horizon: these correspond to the symmetries of two-dimensional Anti de-Sitter space. This space (when projected onto a two-dimensional plane) forms the basis of the M.C. Escher print {\it Circle Limit IV}, which is thus used here to represent the horizon of the black hole. Understanding fermionic response in this system corresponds to scattering bulk fermions off the black hole and shows excitations characteristic of Fermi or non-Fermi liquids.}
\label{fig:apBH}
\end{figure}

Very recently, this apparatus has been brought to bear on the problem
of fermions near quantum criticality~\cite{Lee:2008xf,Liu:2009dm,Cubrovic:2009ye,Faulkner:2009wj}. 
The basic strategy is to perform
angle-resolved photoemission (ARPES)
thought-experiments
on a black hole.
The ground state of a class of strongly-coupled many-body systems
is described by such a charged black hole. The fermionic response,
which is proportional to the ARPES intensity,
may be computed by studying the
scattering of Dirac particles off of this black hole. See fig.~\ref{fig:apBH} for further explanation. In particular, by exploring different regions in parameter space,
both Fermi liquid-like~\cite{Cubrovic:2009ye} and non-Fermi liquid behavior~\cite{Liu:2009dm,Faulkner:2009wj}
were discovered, establishing the black hole as
an important new tool for addressing outstanding questions related to
interacting fermions at finite density.

A prime example of a theoretical challenge
to which such a tool may be applied
is the strange metal phase of the high $T_c$ cuprates. This is a funnel-shaped region in the phase diagram emanating from optimal doping at $T=0$, and
its understanding is believed to be essential for deciphering the mechanism for high $T_c$ superconductivity.
The anomalous behavior of the strange metal---perhaps most prominently the simple and robust linear temperature dependence of the resistivity---has resisted a satisfactory
theoretical explanation for more than 20 years.
In particular it implies that
the low-energy excitations near the Fermi surface are not
Fermi-liquid-like quasiparticles~\cite{varmaetal,anderson}, and suggests that the system may be governed by a quantum critical point at low energies~\cite{aeppli,vallaetal,tallonetl,marel,varma97,varmareview,subir1,subir2}.
Similar anomalous behavior has also been observed in heavy fermion systems near a quantum phase transition~\cite{heavyfermion,janscie,coleman}, where quantum critical behavior is
more clearly established. The anomalous behavior
of a strange metal can be characterized by a phenomenological model called the ``marginal Fermi liquid'' (MFL)~\cite{varmaetal}, which assumes that the spin and charge excitation spectra of the system are momentum-independent (or have weak momentum dependence) and have a specific scale-invariant form.\footnote
{A microscopic model which produces this critical fluctuation spectrum
has been proposed recently in~\cite{morevarma}.}

In this paper we report on a class of non-Fermi liquids discovered using
the AdS/CFT correspondence. The low energy behavior of these non-Fermi liquids is shown to be governed by a nontrivial infrared (IR) fixed point which exhibits nonanalytic scaling behavior only in the temporal direction. This fixed point manifests itself in our dual gravity description as a region containing a two-dimensional anti-de Sitter spacetime. The scaling behavior that arises from it is of the form advocated for the cuprates~\cite{varmaetal,varma97}, and used as an input in the study of heavy fermion criticality~\cite{si:2001}. Within this class of non-Fermi liquids we find examples whose single-particle spectral function and transport behavior resemble those of strange metals.
In particular, the resistivity is proportional to the temperature.
In our treatment, these properties arise from a full quantum-mechanical treatment of the system, and can be understood as being controlled by the scaling dimension of the fermion operator in the emergent IR fixed point. The
fact that the single-particle decay rate and the current dissipation have the same temperature dependence~\cite{varmaetal,varma97} can be understood
holographically in terms of the rate at which Dirac particles fall into the black hole~(see fig.~\ref{Adiss}).

These results offer hope that new insight into the quantum critical points
thought to control the behavior of the cuprates and heavy fermion metals may be gained from a dual gravity description.

\section{Set-up}

 Many examples of field theories with gravity duals are now known in various spacetime dimensions.  Well-studied examples include $\sN=4$ Super-Yang-Mills theory in $d=4$, and ABJM theory in $d=3$~\cite{Bagger:2007vi,Gustavsson:2007vu,Aharony:2008ug}. These theories essentially consist of elementary bosons and fermions interacting with
non-Abelian gauge fields. The rank $N$ of the gauge group is mapped to the Newton
constant $G_N$ of the bulk gravity such that $G_N \propto {1 \ov N^2}$;
the classical gravity approximation in the bulk thus corresponds to the large $N$ limit in the boundary theory. In addition to these theories, there also exist vastly many asymptotically-AdS vacua of string theory~\cite{Douglas:2006es}, each of which is believed to give rise to an example of the correspondence,
though an explicit description of the dual field theory is not known for most vacua.

With a view towards the cuprates, we restrict our discussion to a $(2+1)$-dimensional boundary theory. For simplicity, we take the field theory to be one with a conformally invariant vacuum, which amounts to working with gravity in an asymptotic AdS$_4$ spacetime. This conformal symmetry will not play a role in our results below as it will be broken by putting the system at a finite density.

To set up the gravity description of a system with a finite density of fermions, we consider a boundary theory with a $U(1)$ global symmetry and put the system at a finite chemical potential $\mu$ for the $U(1)$ charge. This can be described in the bulk by a charged black hole whose horizon is topologically $\RR^2$ in AdS$_4$ spacetime~\cite{Chamblin:1999tk}.
 The conserved current $J_\mu$ of the boundary global $U(1)$ is mapped to a bulk $U(1)$ gauge field $A_M$. The black hole is charged under this gauge field, resulting in a nonzero classical background for the electrostatic potential $A_t(r)$. For definiteness we take the charge of the black hole to be positive. The zero temperature limit of the system is described by taking the zero temperature limit of the black hole, which is an extremal
charged black hole. In this paper we will be interested in zero temperature or more generally $T \ll \mu$.
See Appendix~\ref{app:A} for the black hole metric and our normalization of gauge field in gravity action.

\begin{figure}[h]
\begin{center}
\includegraphics[scale=0.50]{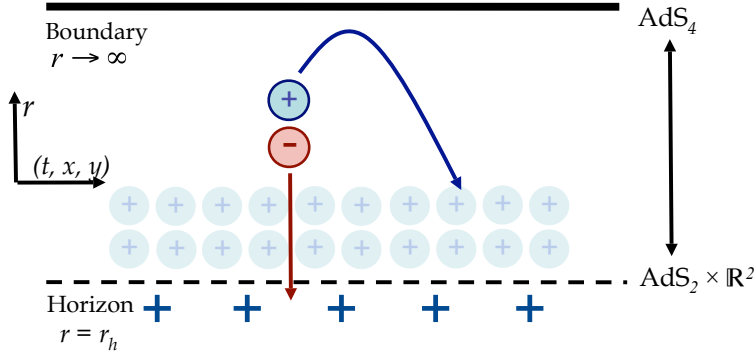}
\end{center}
\vskip -1.cm
\caption{The geometry of an extremal $AdS_{4}$ charged black hole
and the Fermi gas hovering outside it. The horizontal line denotes the boundary spacetime,
which has $d=2+1$ dimensions, and the vertical axis denotes
the radial direction $r$ of the black hole,
which is the direction extra to the boundary spacetime.
The boundary lies at $r=\infty$. The horizon here
is a plane, $\RR^2$.
As will be discussed in section~\ref{sec:IIIA}, the
coordinate $r$ can be interpreted as the energy scale of the boundary theory. The black hole
spacetime asymptotes to that of AdS$_4$ near the boundary and factorizes into AdS$_2 \times \RR^2$ near the horizon, with AdS$_2$ including the $r,t$ directions, and $\RR^2$
comprised by the spatial $x,y$ directions.}
\label{extremalBH}
\end{figure}

Now consider a bulk fermionic field $\psi$. By the general principles of AdS/CFT, this field is dual to some fermionic composite operator $\sO$ in the boundary system. The charge $q$ of $\psi$ under the $U(1)$ gauge field can be identified with global $U(1)$ charge $q$ of $\sO$, and the conformal dimension $\De$ of $\sO$ is related to the mass $m$ of $\psi$ by
 \be
 \De ={3 \ov 2}  + m R \
\ee
where $R$ is the AdS curvature radius. When $(m R)^2 < {q^2 \ov2 }$, quanta for $\psi$ can be pair produced near the horizon~\cite{Pioline:2005pf}. The negative-charged particle in a pair will fall into the black hole, while the positive-charged one moves to the boundary of the spacetime. However it cannot escape, as the curvature of AdS pulls all matter towards its center, and thus the particle will eventually fall back towards the black hole. It then has some probability of falling into the black hole or being scattered
back toward the boundary. This process will eventually reach an equilibrium with a
positively charged
gas of $\psi$ quanta hovering outside the horizon. See figure~\ref{extremalBH}.
In this way we set up a finite density of fermions in the bulk: this
is dual to a finite density of excitations of the fermionic operator $\sO$
in the boundary theory.

The charge density carried by the black hole
is given by the classical geometry and background fields, giving rise to a boundary theory density of order $\rho_0 \sim O(G_N^{-1}) \sim O(N^2)$. In contrast the density for $\psi$ quanta is produced from quantum pair production, giving rise to a much smaller density of fermions of order $\rho_F \sim O(1) \sim O(N^0)$. Thus in the large $N$ limit, we will be studying a small part of a large system, with the background $O(N^2)$ charge density essentially providing a bath for the much smaller $O(1)$ fermionic system we are interested in. Properties of this finite density fermionic system at strong coupling can be found from gravity.\footnote{For other work on this system, see~\cite{reyetal}.} Here we report results on single-particle spectral function and charge transport:

\ben

\item Fermi surface and low energy excitations

Information regarding the fermionic system can be obtained from
the single-particle fermion retarded propagator $G_R (\om, \vk) = \vev{\sO(\om, \vk) \sO(-\om,-\vk)}_{\rm retarded}$ for $\sO$, and the associated spectral function $A (\om, \vk) = {1 \ov \pi} {\rm Im}  G_R (\om, \vk)$. In a physical realization of this system $A(\om,\vk)$ could be measured experimentally by Angular Resolved Photoemission Spectroscopy~(ARPES).
In particular the presence of a Fermi surface and its low energy excitations
manifests itself in the appearance of quasiparticle-like peaks near some shell in momentum space.

\begin{figure}[h]
\begin{center}
\includegraphics[scale=0.50]{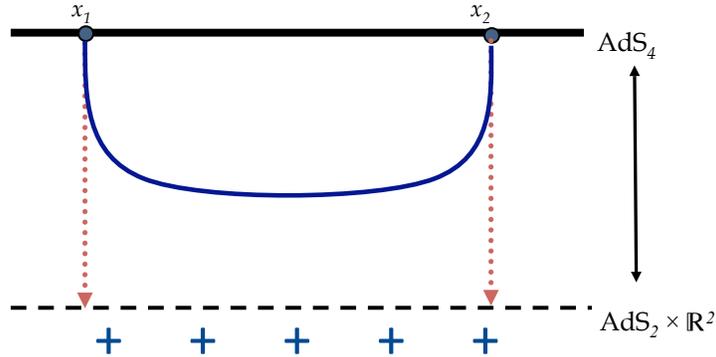}
\end{center}
\vskip -1.0cm
\caption{The two-point function $G_R(x_1, x_2)$ for a fermionic operator $\sO$ in the boundary theory can be obtained from the propagator for the bulk field $\psi$
with endpoints at the boundary. The dissipative part of the correlation function, \ie\ the spectral function $A (\om, \vk) = {\rm Im} \, G_R (\om,\vk)$, is controlled by the rate
at which a $\psi$ quantum falls into the black hole, as indicated by dotted lines in the figure. In particular, in the low frequency limit, this rate is controlled by the geometry of
the near horizon AdS$_2$ region.}
\label{twopF}
\end{figure}

The single-particle retarded function $G_R (\om, \vk)$ for $\sO$ in this strongly coupled system can be obtained from the propagator for the bulk field $\psi$
with endpoints at the boundary~\cite{Iqbal:2009fd,Mueck:1998iz}, as shown in figure~\ref{twopF}. This amounts to solving
the Dirac equation for $\psi$ in the charged AdS black hole
geometry, which takes the form
  \be \label{Dact}
 \Ga^M \sD_M \psi  - m \psi = 0, \qquad
\sD_M  \equiv \p_M + {1 \ov 4} \om_{ab M} \Ga^{ab} - i q A_M
 \ee
where $\Ga^{ab}$ are gamma matrices in the tangent frame, $\om_{ab M}$ is the spin connection for the metric, and there is a nontrivial gauge field $A_M$ due to the charge of the black hole. In the full gravity theory, there are also interaction terms for $\psi$ in~\eqref{Dact}, but they give rise to higher order corrections in $1/N^2$ and can be neglected in the large $N$ limit.

\item Charge transport

\begin{figure}[h]
\begin{center}
\includegraphics[scale=0.50]{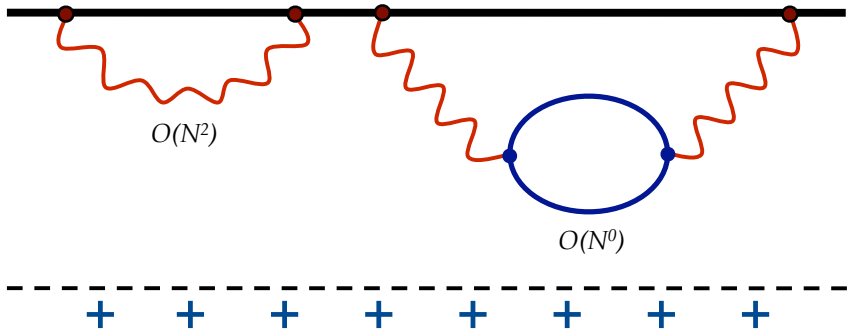}
\end{center}
\vskip -1.0cm
\caption{Conductivity from gravity. The current-current correlator in~\eqref{kubo} can be obtained from the propagator of the gauge field $A_x$ with endpoints on the boundary. Wavy lines correspond to gauge field propagators and the dark line denotes the bulk propagator for
the $\psi$ field. The left diagram is the tree-level propagator for $A_x$, while the right diagram includes the contribution from a loop of $\psi$ quanta. The contribution from the Fermi surface associated with boundary fermionic operator $\sO$ can be extracted from
the diagram on the right.}
\label{twopA}
\end{figure}

The optical conductivity of the fermionic system can be obtained from the Kubo formula \be \label{kubo}
\sig (\om) = {1 \ov i \om} \vev{J_x (\om) J_x (-\om)}_{\rm retarded}
\ee
where $J_x$ is the current density for the global $U(1)$ in $x$ direction at zero spatial
momentum. The right hand side of~\eqref{kubo} can be computed on the gravity side from the propagator of the gauge field $A_x$ with endpoints on the boundary, as in Fig.~\ref{twopA}. In a $1/N^2$ expansion of the conductivity, the leading contribution---of $O(N^2)$---comes from the background black hole geometry. This reflects the presence of the charged bath, and not the fermionic system in which we are interested. As discussed earlier, these fermions have a density of $O(N^0)$, and will give a contribution to $\sig$ of order $O(N^0)$. Thus to isolate their contribution we must perform a one-loop calculation on the gravity side
as indicated in Fig.~\ref{twopA}. Higher loop diagrams can be ignored since they are suppressed by higher powers in $1/N^2$.

\een

To conclude this section we note that our gravity analysis of both single-particle spectral function and conductivity only involve minimal quadratic couplings of $\psi$ to gravity and the $U(1)$ gauge field, which are governed solely by general covariance and gauge symmetry.  That is, they concern with a universal sector of essentially all low energy gravity theories. Thus our discussion applies to a large class of field theories with an AdS gravity dual with a $U(1)$ global symmetry and fermions--the results will depend only on $(q, m)$, and not on the the specific nature of the operator nor the details of the theory.
In particular, we will treat $q$ and $m$ as input parameters, even though in any given theory, only specific values of them will arise. The inclusion of other fields in the bulk
could change the nature of the ground state; we comment further on this issue
in the Discussion section.

\section{Results: Non-Fermi liquids and marginal Fermi liquid }

\subsection{An Infrared fixed point} \label{sec:IIIA}

Before describing the results for the spectral function and conductivities,
we first discuss a feature of the black hole geometry which reveals
a crucial dynamical aspect of the boundary system.
From general principles of AdS/CFT, the coordinate of the extra dimension in the bulk, \ie, the radial direction $r$ of the black hole geometry indicated in figure~\ref{extremalBH}, can be associated with the energy scale in the boundary field theory. This connection, along with the mapping of isometries of the bulk geometry to boundary theory symmetries, provides an organizing principle which associates
geometric aspects of the gravity side to quantum dynamics
of field theory.

The evolution of geometry from the boundary (at $r=\infty$) toward the interior can be interpreted as the renormalization group flow of the boundary theory from high energies (UV) to low energies~(IR).
As $r \to \infty$, the black hole metric approaches
that of AdS$_4$, which has $(2+1)$-dimensional conformal symmetries. This reflects
that at large frequencies $\om \gg \mu$, the effects of the finite density become negligible and one recovers the conformal invariance of the vacuum. For general $r$, the presence of the black hole breaks various symmetries of AdS$_4$, a reflection of the fact that in the boundary theory at scales $\om \sim \mu$, the scaling and Lorentz symmetries are broken by finite density. As one moves close to the
horizon, one finds that the geometry is described by AdS$_2 \times \RR^2$ (see caption of
figure~\ref{extremalBH}). The AdS$_2$ factor, which involves the time and radial direction of the black hole, has an $SL(2,R)$ isometry, \ie\ the symmetry group of conformal quantum mechanics, where the scaling symmetry acts in the time direction.

The presence of the near-horizon AdS$_2$ region and associated symmetries indicates that
at low frequencies (\ie\ $\om \ll \mu$) the boundary system should develop an enhanced symmetry group including scale invariance, and in particular
 should be controlled by a conformal theory, which we will denote the {\it IR CFT} of the boundary theory. This IR CFT only involves the time
 direction, with the spatial directions becoming spectators, an aspect which will be very important below.
  We would like to emphasize that the conformal symmetry of this IR CFT is {\it not} related to the microscopic conformal invariance of the vacuum theory (the UV theory), which at low frequencies is completely broken by the finite charge density. This new infrared conformal symmetry apparently emerges as a consequence of the collective behavior of a large number of degrees of freedom.  

It is currently not known how to describe this IR CFT using the conventional language of a Lagrangian. Its properties, however, are fully computable from the AdS$_2$ gravity dual.
For example, the form of the bulk spinor wave equation implies the following.
The fermion operator with momentum $\vk$, $\sO(\vk)$,
matches onto an operator $\sO_k^{IR}$ in the IR CFT
with an anomalous scaling dimension $\delta_k$ given by
 \be \label{scaldim}
\delta_k  =\ha + \nu_k , \qquad \nu_k  = {1 \ov \sqrt{6}} \sqrt{{m^2 R^2} +  {3 k^2 \ov \mu^2} - {q^2 \ov 2}}, \qquad k = |\vk| \ .
\ee
Furthermore one can compute the two-point correlation functions of $\sO_k^{IR}$ in the IR CFT, which we denote by $\sG_k(\om)$. The emergent conformal invariance fixes these correlation functions to be of the form
\be \label{iRc}
\sG_k(\omega) = c(k)
\om^{2 \nu_{k}},
\ee
where the prefactor $c(k)$ is complex and analytic in momentum $k$.
It is not fixed by conformal invariance and is given explicitly in Appendix~\ref{app:A}.
Note that due to spherical symmetry both $\nu_k$ and $\sG_k$ depend only on the modulus $k$ of $\vk$.
 It is important to note that these correlation functions $\sG_k(\om)$ are not the correlation functions of the full theory, which will be described in the next section: however, we will see that they directly control the small-frequency dependence of the full spectral densities.

Since we are working with a theory which is scale invariant in the vacuum, all dimensional quantities can be expressed in units of the chemical potential $\mu$. In the following, to simplify notations we often do not write factors of $\mu$ explicitly, which can be reinstated from simple dimensional analysis.

\subsection{One-particle spectral function}

The retarded function $G_R (\om, \vk)$ for $\sO$ in the full theory is found by solving Dirac equation~\eqref{Dact} in the full bulk black hole geometry. Here we summarize the result. The details are rather involved and are reported elsewhere~\cite{Liu:2009dm,Faulkner:2009wj}.

Due to spherical symmetry, $G_R$ only depends on $k=|\vk|$, up to a change of basis.
We first describe its properties at zero temperature. In the low frequency limit $\om \ll \mu$, where $\mu$ is the chemical potential, it is possible to express $G_R (\om, k)$ in terms of the correlation function $\sG_k (\om)$ for $\sO (\vk)$ in the near horizon AdS$_2$ region plus quantities which depend on the full geometry, to which we will refer as UV quantities to distinguish them from objects such as~\eqref{scaldim} and~\eqref{iRc} that depend only on the structure of the IR CFT. At a generic value of $k$, we find a low frequency expansion
\be
G_{R} (\om, k) = F_0 (k) + F_1 (k) \omega +
F_2 (k) \sG_k(\omega)+ \dots
\ee
where UV quantities $F_{0,1,2} (k)$ are real and analytic functions of $k$ and $\sG_k (\om)$ is given by~\eqref{iRc}. The imaginary part of the correlator and thus the spectral function is controlled by $\sG_k (\om)$.
This has a simple geometric interpretation as indicated in figure~\ref{twopF}.

For $m^2 R^2 < {q^2 \ov 3}$, the Dirac equation~\eqref{Dact} can have a static normalizable solution at some discrete shell in momentum space, which we call $|\vec k| = k_F$. Near such a special value of $k$, which can be determined numerically, $G_R$ takes a very different form and its small-frequency expansion can be written as
\def\hh{h}
\be \label{thrIM}
 G_{R} (k,\om)
  =
{h_1 \ov  k-k_F -  {1 \ov v_F} \om -  \Sig (\om,k)} , \qquad \Sig (\om,k) = \hh \sG_{k_F}(\omega)=  \hh c(k_F)\om^{2 \nu_{k_F}}
 \ee
where $\sG_{k_F} (\om)$ is given by~\eqref{iRc} evaluated at $k=k_F$.
The quantities $v_F, h_1$ and $h$ in~\eqref{thrIM} are positive constants
which are known numerically. The associated spectral function
$A (\om, k) ={1 \ov \pi} {\rm Im}\,
G_R (\om, k)$ is given by
 \be \label{specT}
 A (\om,k) = {1 \ov \pi} {h_1 \Sig_2 \ov (k-k_F - {1 \ov v_F} \om - \Sig_1)^2 + \Sig_2^2}, \qquad
 \Sig (\om, k) \equiv \Sig_1 (\om,k) + i \Sig_2 (\om,k) \ .
 \ee
The non-analytic frequency dependence of
the spectral density $A (\om,k)$ is again controlled by the IR CFT correlation function.

\begin{figure}[h!]
 \begin{center}
\includegraphics[scale=0.85,angle=-90]{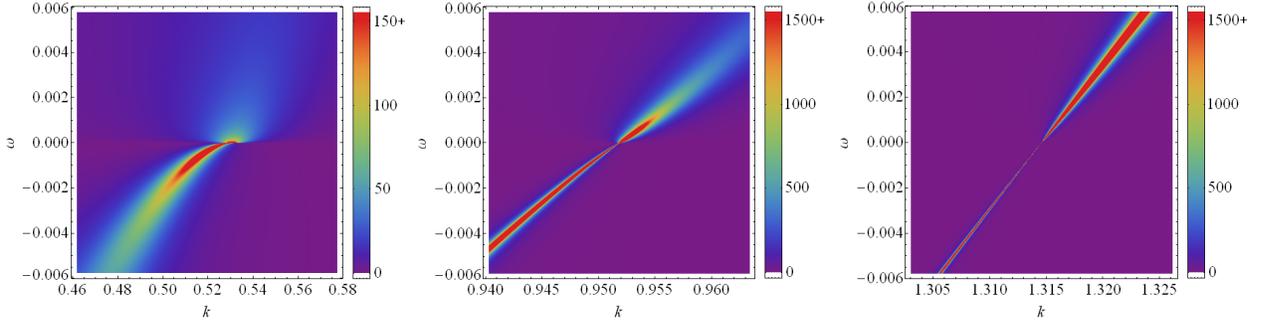}
\caption{\label{fig:fits}
Density plot of the spectral function $A (\om,k)$ as a function of $\omega$ and $k$.
We have chosen units so that $\mu = 1$ and in all plots $m=0$.
{\it Left}: $q=1$ with Fermi momentum $k_F= 0.53$ and $\nu_{k_F} = 0.24 < \ha$ for which there is no sharp quasiparticle at the Fermi surface and the peak dispersion
is nonlinear (see equation~\eqref{peaD}). {\it Middle}: $q = 1.56$ with
Fermi momentum $k_F = 0.952$ and $\nu_{k_F} = 0.500$ which corresponds to a marginal Fermi liquid. {\it Right}: $q=2$ with Fermi momentum $k_F = 1.315$ and $\nu_{k_F}   = 0.73 > \ha$, for which there are stable quasi-particles at the Fermi surface.
}
\end{center}
\end{figure}

Equation~\eqref{thrIM} has a pole in the lower frequency plane which approaches $\om =0$ as $k \to k_F$. As a result, the associated spectral function
has a quasiparticle-like peak whose location as a function of $\om$ approaches
$\om =0$ for $k \to k_F$. The height and width of the peak approach infinity and zero respectively. We take this as evidence that the
system has a Fermi surface at $k= k_F$. The quantity $\Sig (k,\om)$, which only depends on $\om$ (up to higher order analytic corrections in $k-k_F$), can be interpreted as the self-energy
for excitations near the Fermi surface, and its imaginary part $\Sig_2$ can be interpreted as the single-particle scattering rate near the Fermi surface.
See fig.~\ref{fig:fits} for a density plot of $A (\om,k)$ as a function of
$k$ and $\om$.

The two frequency-dependent terms in the downstairs of~\eqref{thrIM}
have distinct physical origins. The term linear in $\om$ comes from an analytic
expansion of UV quantities and is real, while the self-energy $\Sig (\om)$ comes from the near horizon AdS$_2$ region and is complex, scaling with frequency like $\om^{2\nu_{k_F}}$.
The properties of the quasiparticle-like peak depend on the competition between these two terms. The term linear in $\om$ in the denominator of~\eqref{thrIM} can be omitted if $\nu_{k_F} < \ha$. For $\nu_{k_F} > \ha$ we should still keep the term proportional to $\om^{2 \nu_{k_F}}$, since it makes the leading contribution to the imaginary part.
When $\nu_{k_F}<\ha$, equation~\eqref{thrIM} has a scaling form consistent with
the scaling hypothesis for a critical Fermi surface discussed recently by Senthil~\cite{senthil:0803}, while for $\nu_{k_F} > \ha$ it has
stable quasi-particles similar to a Fermi liquid.

More explicitly, when $\nu_{k_F} > \ha$,
the Green's function in equation~\eqref{thrIM} has a pole in the lower half complex
$\om$-plane located at
 \be
 \om_c (k) \equiv  \om_* (k) - i \Ga (k)
 \ee
with
\be
\om_* (k) = v_F (k-k_F) + \cdots, \qquad {\Ga (k) \ov \om_* (k)}
\sim (k-k_F)^{2 \nu_{k_F} -1} \to 0, \qquad
\ee
and the residue of the pole is $Z = h_1 v_F$. The pole represents a quasiparticle which has a linear dispersion with $v_F$ as the (Fermi) velocity. It becomes stable approaching the Fermi surface, with a non-vanishing spectral weight $Z$ at the Fermi surface itself.
In this case the standard quasiparticle picture applies even though the scaling exponent of the scattering rate is generically different from the $\omega^2$-dependence of the Landau Fermi liquid.

For $\nu_{k_F} < \ha$, one instead finds that as $k$ is varied toward $k_F$
 \be \label{peaD}
 \om_* (k) \sim (k-k_F)^{z}, \qquad z={1 \ov 2 \nu_{k_F}} > 1, \qquad
 {\Ga (k) \ov \om_* (k)} = {\rm const}
 \ee
Also the residue at the pole approaches zero at the Fermi surface
\be
Z \propto (k-k_F)^{1- 2 \nu_{k_F} \ov 2 \nu_{k_F}} \to 0 \ .
\ee
In this case the imaginary part is always comparable to the real part
and the quasiparticle is never stable. Also note that the residue of the pole (quasiparticle weight) vanishes at the Fermi surface. This is thus an example of a
Fermi surface without sharp quasiparticles!

When $\nu_k = \ha$, the two frequency-dependent terms in~\eqref{thrIM}
become degenerate in the small frequency limit. As is often the case for
such a degenerate situation, a logarithmic term is generated, with the self-energy given by
 \be \label{eprM}
 \Sig (\om) \approx  \tilde c_1 \om \log \om + i d_1 \om, \quad {d_1 \ov \tilde c_1}
 = -{\pi \ov 1 + e^{- {2 \pi q \ov \sqrt{12}}}}
  \ee
where $\tilde c_1 < 0$ and $d_1$ are {\it real} constants which can be obtained from~\eqref{defck}.
Thus in this case the single-particle scattering rate is {\it linear} in $\om$. While it is still suppressed compared to the real part as the Fermi surface is approached, but the suppression is only logarithmic.
The quasiparticle residue also vanishes logarithmically at the Fermi surface. Remarkably~\eqref{eprM} is of the form\footnote{Note for marginal Fermi liquids proposed in~\cite{varmaetal},  ${d_1 \ov \tilde c_1}$ is equal to $-{\pi \ov 2}$ for which case there is a particle-hole symmetry, i.e. $\Sig_2 (-\om) = \Sig_2 (\om)$. In~\eqref{eprM}
this happens only for $q=0$. The particle-hole asymmetry in our case for generic $q$ can be attributed to the fact that quantum critical fluctuations are charged. Also note in our case~\eqref{eprM} receives analytic corrections in $k$ in both exponent and prefactors.} postulated in~\cite{varmaetal} for the ``Marginal Fermi Liquid'', which fits well with results from ARPES experiments on cuprates in the strange metal region~\cite{abrahams2000}.

Similar logarithmic terms appear
for any $\nu_{k_F} = {n \ov 2}, \;\; n \in \ZZ_+$. For example, at $\nu_{k_F} =1$, one finds that the self-energy becomes
 \be \label{jejr}
 \Sig (\om) \approx  \tilde c_2 \om^2 \log \om + c_2 \om^2
 \ee
with $\tilde c_2$ real and $c_2$ complex constants. Equation~\eqref{jejr} resembles that of a Landau Fermi liquid with the imaginary part of the self-energy quadratic in $\om^2$, but it has a logarithmic term $ \om^2 \log \om$ with a real coefficient which is not present for a Landau Fermi liquid.

At a nonzero temperature $T \ll \mu$, one finds the self-energy $\Sig (\om)$ in equation~\eqref{thrIM} is replaced by
\be\label{finiteTspinorG}
 \Sig (\om, T) =   h_2 \sG_{k_F}(\omega, T)
 \ee
where $\sG_{k_F}(\omega, T) = (2 \pi T)^{2 \nu_{k_F}} g_1\( {\omega \over T}, {k_F \over \mu} \)$
is the finite-temperature IR CFT Green's function. $g_1$ is a universal scaling function which is given explicitly in Appendix~\ref{app:A}. It approaches a constant for small $\om/T$ and gives the correlator at zero temperature in the limit $\om/T \to \infty$.
At a finite temperature, the pole of~\eqref{thrIM} always has a negative nonzero imaginary part and never reaches the real axis. This reflects the familiar fact that Fermi surface
gets smeared at finite temperature. For the Marginal Fermi Liquid case, $\nu_{k_F} = \half$, the self-energy is $\Sig (\om) = \pi T g_2 (u)$ with $u \equiv {\omega \ov 2 \pi T} - {q \ov \sqrt{12}}$ and $g_2 (u)$ is a scaling function given by
\be \label{marF}
g_2 (u) =  2 i d_1 u  + \tilde{c}_1 \le(2 u \log{T \ov \mu} + 2u
 \psi( - i u) + i \pi u + i \ri)  + \cdots
\ee
where  $\psi$ is the digamma function, $\tilde c_1$ and $d_1$ are the same constants as in~\eqref{eprM}, and $\cdots$ denotes terms which are real and analytic in $\om$ and $T$.

\subsection{DC and optical conductivities}

\begin{figure}[h]
\begin{center}
\includegraphics[scale=0.50]{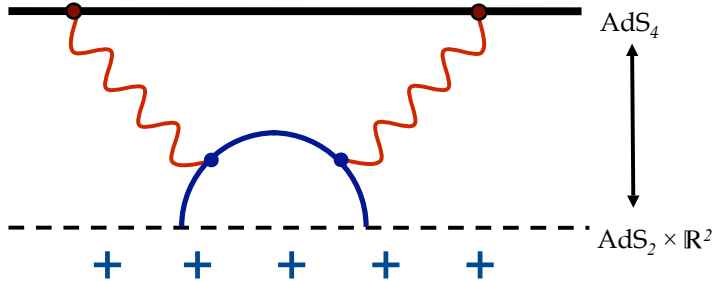}
\end{center}
\vskip -1.0cm
\caption{The imaginary part of the current-current correlator~\eqref{kubo} receives its dominant contribution from diagrams in which the fermion loop goes into the horizon. This also gives an intuitive picture that the dissipation of current is controlled by the decay
of the particles running in the loop, which in the bulk occurs by falling into the black hole.
}
\label{Adiss}
\end{figure}

The contribution of the Fermi surface to the conductivity can be extracted
from the one-loop diagram in figure~\ref{twopA}, whose evaluation proceeds similarly to
that of a Fermi liquid, except for the extra bulk dimension and complications from curved spacetime geometry. There is also an important qualitative difference: we are evaluating the diagram in a spacetime with an event horizon. For example, the fermion running in the loop can go into the horizon as indicated in Fig.~\ref{Adiss}. In fact,
precisely such processes give the leading contribution to the dissipative part of the current-current correlator~\eqref{kubo}, and thus the optical and DC conductivities.

Since 
the system contains a charged background $O(N^2)$ fluid, the DC conductivity at zero temperature is not really well-defined as there is no mechanism to dissipate away the
current. In a sense,  the system should behave like a perfect conductor. The signature
of this in the optical conductivity is a delta function at zero frequency with
a coefficient of $O(N^2)$~\cite{Hartnoll:2007ih}.
We will instead be interested
in the leading low temperature contribution from the Fermi surface to the conductivity. A well-defined answer for DC conductivity can be extracted since at any nonzero temperature of $O(N^0)$, the system contains, along with the charged background fluid,
a neutral component with density of $O(N^2)$ which can dissipate away the $O(N^0)$ momenta of the fermionic quanta\footnote{The optical conductivity  should also contain a piece proportional to $\delta (\om)$ at $O(N^0)$ level, which can be interpreted as a correction to the corresponding $O(N^2)$ piece. The contribution from the Fermi surface does not give rise to a delta function at any non-zero temperature.}.

\begin{figure}[h]
\begin{center}
\includegraphics[scale=0.35]{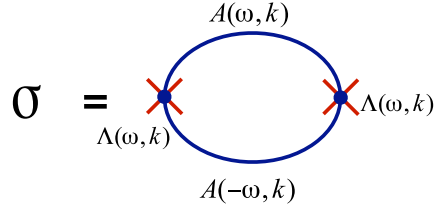}
\end{center}
\vskip -1.0cm
\caption{Equation~\eqref{eqn:foet} can be interpreted in terms of a Fermi liquid picture with the exact propagator for a fermion running in the loop and an effective vertex
$\Lam$.}
\label{bubble}
\end{figure}

We find the conductivity~\eqref{kubo} can be written in terms of boundary theory quantities as
\be
 \sig(\omega)
 =   {C \ov  i \omega }  \int \! d \vk \int {d \om_1 \ov 2 \pi}
 {d \om_2 \ov 2 \pi}  \,
  { f(\omega_1) - f(\omega_2) \over \om_1 - \om -\om_2 - i \ep} 
   A (\om_1 ,\vk) \,  \vertexZ(\om_1, \omega_2, \omega, \vk)
  \vertexZ (\om_2, \omega_1, \omega, \vk)
 \, A (\om_2 ,\vk) \,
 \label{eqn:foet}
 \ee
where $f(\om) = {1 \ov e^{\om \ov T} +1}$ is the Fermi distribution function, $A (\om,\vk)$ is the single particle spectral function discussed in last subsection, $\Lam$ is an effective vertex which can be evaluated explicitly, and $C$ is a temperature independent overall constant. The explicit derivation and evaluation of~\cite{conductivityref} are rather complicated and details will appear elsewhere~\cite{conductivityref}. For an outline of the derivation see Appendix~\ref{app:A}, where an expression for $\Lam$ is also given. Equation~\eqref{eqn:foet} can be interpreted
in the boundary theory, as seen in Fig.~\ref{bubble}, in terms of standard Fermi liquid
theory where $\Lam$ plays the role of an effective vertex.
$\Lam$ is in general a complicated function of $\om, \vk$ and $T$, but
in the low temperature limit and near the Fermi surface it becomes a smooth, real function of the modulus $k$, independent of $\omega$ and $T$. In this limit, the conductivity is controlled by the single-particle spectral function near the Fermi surface.

We find the DC conductivity, given by $\om \to 0$ limit of~\eqref{eqn:foet}, can be written as
 \be \label{sigS}
 \sig_{DC} = \al (q,m) \, T^{-2 \nu_{k_F}}  \
 \ee
with $\al (q,m)$ a numerical constant depending on the input parameters $q$ and $m$.
In particular, for $\nu_{k_F} = \ha$, which corresponds to the Marginal Fermi Liquid case, we find the contribution to the resistivity from the Fermi surface is {\it linear}, as observed for cuprates in the strange metal phase and many other non-Fermi liquid materials~\cite{stewart}.

In general, the electrical transport scattering rate can have different temperature and frequency dependence from the single-particle scattering rate, because the former emphasizes large momentum transfer. Here we find that the scaling exponent in~\eqref{sigS} is the same as that for $\Sig$ in~\eqref{thrIM} and~\eqref{specT}.
On the gravity side, this can be understood intuitively from figure~\ref{Adiss}, which indicates that the dissipative part of the
current-current correlator is controlled by the rate of the bulk fermion falling through the horizon, which as described earlier also gives the single-particle scattering rate.
From the boundary theory point of view, this is possible because the single-particle decay
rate is controlled by the IR CFT, which
is critical only in the time direction and does not
distinguish between small and large momentum transfer.

The optical conductivity has the advantage that it can distinguish situations with or without quasiparticles, since the existence of a quasiparticle gives rise to a scale which is associated with the lifetime of the quasiparticle and is manifest through a transport peak
at $\om =0$. We will be interested in the regime $\om , T \ll \mu$ with $\om/T$ arbitrary.
More explicitly, we find that for $\nu_{k_F}<1/2$, where there are {\it no} stable quasiparticles,
\be \label{ss1}
\sigma (\om) = T^{-2\nu_{k_F}}  F_1( \omega/T)\,.
\ee
Here $F_1 (x) $ is a universal scaling function approaching a constant as $x \to 0$. For large $x$ it falls off as $x^{-2 \nu_{k_F}}$ which leads to a conductivity of the form,
\be
\label{fall}
\sig (\om) = a (i \om)^{-2 \nu_{k_F}}, \qquad T \ll \om \ll \mu \
\ee
with $a$ a real constant. Note that the phase factor above is fixed by time reversal symmetry which requires $\sig^* (\om) = \sig (-\om)$. The falloff in (\ref{fall}) is much slower than the Lorentzian form
familiar from Drude theory. The behavior~(\ref{ss1}-\ref{fall}) is consistent with that of a system without a scale and with no quasiparticles. 

For $ \nu_{k_F}> 1/2$, there are two regimes. In the first we
take $u \equiv {\om  T^{-2 \nu_{k_f}}}$ fixed in the low temperature limit. Then one finds that $\sig$ can be approximated by a Drude form
\be \label{drude}
\sig (\om) = {\om_p^2 \ov {1 \ov \tau} - i \om}, \quad {\rm with} \quad \om_p^2 = {v_F C' \ov 2 \pi}, \qquad {1 \ov \tau} = 2 v_F \Im \Sig (\om=0) \propto T^{2 \nu_{k_F}} \ll T
\ee
where $C'$ is introduced in~\eqref{cond} in Appendix~\ref{app:A} and $v_F$ is the Fermi velocity. As commented before the transport scattering rate $1/\tau$  is proportional to the single-particle scattering rate $\Im \Sig$. When $\mu \gg \om \gg T$, we find the scaling
behavior
\be \label{larfe}
\sig (\om) ={i \om_p^2 \ov \om}+  b (i \om)^{2 \nu_{k_{F}}-2}
\ee
with $b$ a real constant. The $1/\om$ term gives rise to a term proportional to $\delta (\om)$ with a weight consistent with that in~\eqref{drude}.

 For $\nu_{k_F} = \ha$, the Marginal Fermi liquid, we find
 \be
 \sig(\om) = T^{-1} F_2 \le({\om \ov T}, \log {T \ov \mu} \ri)
 \ee
where $F_2$ approaches a $T$-independent constant at small $\om$. Due to time
reversal symmetry the real part $\sig_1 (\om)$ of $\sig (\om) $ is an even function in $\om$ and thus for $\om/T < 1$, one can again approximate $\sig (\om)$ by a Drude form with
the transport scattering time $\tau \propto {1 \ov T}$. For $\om \gg T$ we find that
 \be
 \label{eq:marginaloptical}
 \sig (\om) \sim {1 \ov \om} {i C'\ov 2   \tilde c_1}  \le({1 \ov   \log {\om \ov \mu}}
 + {1 \ov  (\log {\om \ov \mu})^2} {1 + i \pi \ov 2 } \ri) + \cdots
 \ee
where $\tilde c_1$ introduced in~\eqref{eprM} and~\eqref{marF} is negative. The above expression is consistent with that recently obtained for marginal Fermi liquids from
summation of the particle-hole ladder in~\cite{varmahydro}.

In the strange metal region of the phase diagram for cuprates, it was observed in~\cite{Azraketal,marel} that the optical conductivity for Bi-2212 has two different
scaling regions: (i) for $\om < T$, it has a Drude form  with ${1 \ov \tau} \sim T$
reflecting temperature as the only scale, and (ii) for $\om > T$,
$\sig (\om) = C(i \om)^{\ga -2}$ with $\ga \approx 1.33$.  In region (i) the strange metal appears to be consistent with the Marginal Fermi Liquid behavior with $\nu_{k_F} =\ha$. But interestingly in region (ii) the strange metal resembles the behavior~\eqref{larfe} for $\nu_{k_F} = {2 \ov 3} > \ha$ and appears to be consistent with that proposed by
Anderson~\cite{Anderson} based on theories of Luttinger liquids.

\section{Discussion}

\begin{figure}[h]
\begin{center}
\includegraphics[scale=0.40]{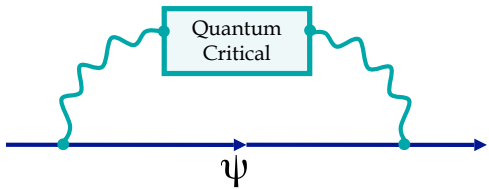}
\end{center}
\vskip -1.0cm
\caption{A cartoon of the underlying physical mechanism of our non-Fermi liquids~\cite{Faulkner:2009wj,Faulkner:2010tq}. }
\label{conf}
\end{figure}

The physical picture underlying our non-Fermi liquids can be summarized as in figure~\ref{conf}. Quantum critical fluctuations indicated in the
figure are described by degrees of freedom in the IR CFT and are represented
 in the gravity side by the near horizon AdS$_2$ region. An interesting feature here is that critical fluctuations remain critical for any momentum, similar to that
 postulated in the Marginal Fermi Liquid description of the cuprates.
Here it comes from first-principle computations with a full quantum-mechanical
treatment of the system.

Non-Fermi liquids also arise from coupling a Fermi liquid to a gapless propagating bosonic mode, such as a transverse magnetic excitation of a gauge field,~\eg\cite{Holstein:1973zz,patrick}. 
The forms of the fermion Green's functions thus obtained fit into the set of functions we have found in~\eqref{thrIM}.
An important difference, however, is that for a gapless boson small momentum scatterings are strongly preferred due to its linear dispersion relation.
As a result the corresponding resistivity grows with temperature as a higher power than the single-particle scattering rate~\cite{patrick}.

The extremal black hole we are studying has a `residual' zero-temperature entropy.
This degeneracy is exact in the $N\to\infty$ limit; at finite $N$ it could be lifted to a low-lying density of states, as in a system with frustration.
Nevertheless it is interesting to ask whether the IR CFT and its properties are tied to this zero temperature entropy. Such a correlation may give a hint about the nature of quantum critical points observed in realistic condensed matter systems~\cite{zaanen:jc}.
Recently it has been argued in~\cite{Hartnoll:2009ns} that when the fermionic backreaction to the black hole geometry is included, the black hole horizon is replaced by a geometry with no zero-temperature entropy.  In the large $N$ limit corrections due this replacement are higher order in $N^2$~\cite{Faulkner:2010tq}
and qualitative features of our earlier discussion are not affected.

In a full gravity theory where the extremal charged black hole arises, there could also be light charged scalar fields. A mechanism similar to the one described earlier which leads to the bulk Fermi gas in Fig.~\ref{extremalBH} leads to condensation of those scalar fields, and a superconducting state in the boundary theory~\cite{holographicsc}.
Thus the Fermi surface state and the IR CFT we are describing could be hidden behind a superconducting state. In fact, there are indications that the existence of such light charged scalars may be generic in string theory compactifications~\cite{Denef:2009tp}.
This is not dissimilar to many known condensed matter systems where
quantum critical points appear to have a tendency to be hidden under some superconducting dome. This resemblance may not be an accident and deserves further study.

The effect of such a condensate on the Fermi surfaces discussed here has been
explored in \cite{Faulkner:2009am, chenkaowen, fabio}.
In such a superconducting state
a well defined gapped quasiparticle appears
for a reasonable coupling between the bulk scalar and spinor \cite{Faulkner:2009am}.
For energies above the gap the behavior is as discussed
here -- an IR CFT is still controlling the physics of the quasiparticles despite
the presence of the condensate. One can get rid of superconducting instability
by turning on a magnetic field, which has been discussed recently in~\cite{Albash:2009wz,Basu:2009qz,denefetal,Hartnoll:2009kk},
and quantum oscillations in the Fermi surface contribution to the free energy have
been observed in~\cite{denefetal,Hartnoll:2009kk}.

To conclude, our models provide a laboratory for studying qualitative features of non-Fermi liquids, and offer hope that new insights into the quantum critical points
thought to control the behavior of
the high $T_c$ cuprates and heavy fermion metals may be gained from a dual gravity description. For example, the emergent IR fixed point
and its consequences for single-particle particle and transport may transcend the particular family of gravity models that we are studying, and may have applications to real systems.

\appendix

\section{Supplementary materials} \label{app:A}

\subsection{Gravity action and black hole metric}

The action for a vector field $A_M$ coupled to AdS$_{4}$ gravity can be written as
 \be \label{grac}
 S = {1 \ov 2 \kappa^2} \int d^{4} x \,
 \sqrt{-g} \le[\sR +  { 6 \ov R^2} - R^2  F_{MN} F^{MN} \ri]
\ee
and $R$ is the
curvature radius of AdS. The equations of motion following from~\eqref{grac} are solved by the geometry of a charged black hole~\cite{Romans:1991nq,Chamblin:1999tk}
 \be \label{bhmetric1}
 {ds^2 \ov R^2} \equiv g_{MN} dx^M dx^N =  {r^2 } (-f dt^2 + d\vec x^2)  +  {dr^2 \ov r^2 f}
 \ee
 with
 \be \label{bhga2}
 f = 1 + { \mu^2 \ov r^{4}} - {1+\mu^2 \ov r^3}, \qquad A_t = \mu \le(1- {1 \ov  r}\ri), \qquad  \
 \ee
where  $\mu$ is the chemical potential in dimensionless units. The horizon is at $r =1$ and the Hawking temperature is $T = {3-\mu^2 \ov 4 \pi}$.

\subsection{Determination of $k_F$ and $\nu_{k_F}$ dependence on $q,m$}

\begin{figure}[h]
\begin{center}
\includegraphics[scale=0.45]{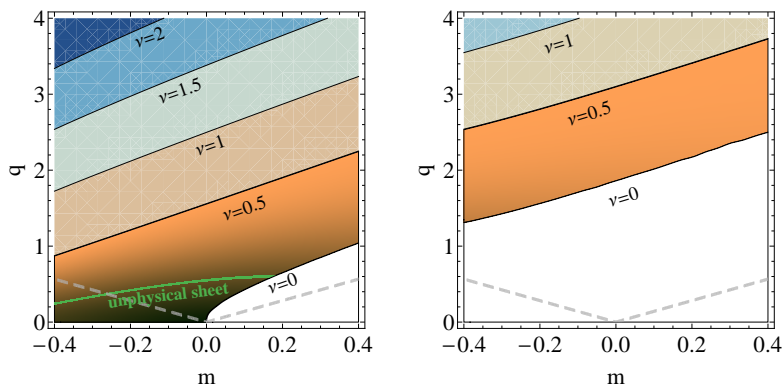}
\end{center}
\vskip -1.0cm
\caption{Distribution of $\nu_{k_F}$ in the $m-q$ plane. The two plots correspond to two components for a spinor operator in $(2+1)$-dimension, which we have been suppressing
in the main text. In the white
region, there is no Fermi surface. There is no quasiparticle in the orange region, $\nu_{k_F} < \ha$. In the
remainder of the parameter space, there is a stable quasiparticle.}
\label{phase}
\end{figure}

A Fermi surface corresponds to a normalizable mode of the Dirac equation in the black hole background. Because of the presence of the horizon, this can only occur for static solutions, $\omega =0$. Squaring the static Dirac equation and rewriting it as a Schr\"{o}dinger equation one can then reduce the problem of finding a Fermi surface to finding bound states in the resulting potential as a function of $k$. Bound states appear for discrete values of $k = k_F$. Depending on parameters $(q,m)$ there can be multiple bound states which corresponds to the system having multiple Fermi surfaces. For any nonzero frequency the Scro\"{o}dinger potential becomes unbounded at the horizon and the state can tunnel through a barrier into the horizon. Estimating the decay rate using WKB one finds the width of the state is roughly $\max(|\omega|^{2\nu_{k_F}}, T^{2\nu_{k_F}})$ consistent
with the self energy of the quasi-particle in the field theory. With $k_F$ determined, the scaling exponent $\nu_{k_F}$ can then be computed from~\eqref{scaldim}. The result is presented in fig.~\ref{phase}.

\subsection{Some long expressions}

The prefactor $c(k)$ in equation~\eqref{iRc} is given by
\be \label{defck}
c(k) =  e^{- i \pi \nu_k} \frac{\Gamma (-2 \nu_k ) \, \Gamma \left(1+\nu_k -i {q \ov \sqrt{12}} \right)}{\Gamma (2 \nu_k )\,   \Gamma \left(1-\nu_k -i {q \ov \sqrt{12}}\right)}
 ~\frac{ \le({mR \ov \sqrt{6}}  + i {k \ov \sqrt{2} \mu}  \ri)- i {q \ov \sqrt{12}} - \nu_k}
{ \le({mR \ov \sqrt{6}}  + i {k \ov \sqrt{2} \mu}  \ri) - i {q \ov \sqrt{12}} +  \nu_k},
\ee
This expression is reminiscent of those for certain two-dimensional CFT. It has been argued in~\cite{Guica:2008mu,Lu:2009gj} that the asymptotic symmetry group of the near horizon AdS$_2$ region is generated by a single copy of Virasoro algebra with a nontrivial central charge. 
The scaling function $g_1$ introduced below equation~\eqref{finiteTspinorG}
can be written as
\be
 g_1\( {\omega \over T}, {k \over \mu} \) =  e^{i \pi \nu_k} c(k) {\Gamma (\ha+\nu_k -\frac{i \omega }{2 \pi T }+i {q \ov \sqrt{12}} ) \ov \Gamma \left(\frac{1}{2}-\nu_k -\frac{i \omega }{2 \pi T}+i
{q \ov \sqrt{12}} \right)}
\ee
where $c(k)$ is given by~\eqref{defck}.

\subsection{Computation of conductivity}

Here we outline the computation of conductivity~\eqref{eqn:foet}. Details will appear in~\cite{conductivityref}.
Our strategy is to first write down an integral expression for the two-point
current correlation function $G_E^{(j)} (i \om_l)$ in Euclidean signature and then analytically continue to Lorentzian signature inside the integral (for  $\om_l > 0$)
  \be \label{anaLy}
 G_R^{(j)} (\om) = G_E^{(j)} (i\om_l = \om + i \ep) \
 \ee
where $G_R^{(j)} (\om)$ denotes the Lorentzian current correlation function on the right hand side of~\eqref{kubo}. The important steps in obtaining $G_E^{(j)}  (i \om_l)$ are:

\ben

\item In a charged black hole geometry, fluctuations of the $U(1)$ vector field $A_x$ which is dual to the boundary current $J_x$ mixes with metric fluctuations $h_{tx}$. Thus the wiggly lines in fig.~\ref{twopA} contain the boundary-to-bulk propagators for both vector and metric fields.

\item In addition to cubic couplings (schematically) $\bar \psi  \psi A$ and $\bar \psi \psi  h $ indicated in fig.~\ref{twopA}, the couplings of
    internal fermions $\psi$ to vector and metric fields also include quartic couplings like $h^2 \bar \psi \psi $ and $h A \bar \psi \psi$, which give rise
to seagull-like diagrams. It can be shown by explicit calculation that the segull-like
diagrams are always subleading.

\item Bulk-to-bulk spinor propagator $D_E (r_1,r_2;i \om_m,\vk)$ for $\psi$ as indicated in solid lines in fig.~\ref{twopA} can be written using its spectral decomposition
     \be
D_E(r_1,r_2;i\om_m,\vk) = \int \frac{d{\omega}}{2\pi} \frac{\rho(r_1,r_2;\omega,\vk) }{i\om_m - \omega},
\label{spectrep}
\ee
where $\rho(r_1,r_2;\omega,\vk)$ is the bulk spectral function for $\psi$, where we have suppressed bulk spinor indices. The bulk spectral function $\rho$ can in turn be related to the boundary spectral function $A_{\al \beta} (\om,\vk)$~\eqref{specT} as (we have restored the boundary spinor indices $\al, \beta$):
  \begin{equation}
  \label{specBB}
\rho(r_1,r_2;\omega,\vk) = \psinorm_{\alpha}(r_1;\om,\vk) \, A_{\alpha \beta} (\om,\vk)\, \overline{\psinorm_{\beta}} (r_2;\om,\vk) \ .
\end{equation}
$\psinorm_\al$ is a normalizable {\it Lorentzian} bulk wave function for $\psi$. Again the bulk spinor indices are suppressed. The boundary spinor indices $\al, \beta$ on $\psinorm$ label independent bulk solutions.

\een
With the above ingredients and analytic continuation~\eqref{anaLy}, evaluating the one-loop diagram in fig.~\ref{twopA} one finds~\eqref{eqn:foet} with the effective vertex $\Lam$ given by
\be \label{effV}
 \vertexZ_{\beta \gamma} (\om_1, \om_2, \om,\vk) = \int dr \sqrt{-g} \, \overline{\psinorm_\beta} (r; \om_1,\vk) \, Q (r;  \om; \vk) \, \psinorm_\gamma (r; \om_2, \vk)
 \ee
with ($K_h$ and $K_A$ are the boundary to bulk propagators for $h_{tx}$ and $A_x$ respectively)
\be \label{kern1}
 Q(r; \om, \vk) =- i \le(- i  k_x  K_h (r;\om) \Ga^{t} + {r^2 \ov 8}   \p_r K_h (r;\om) \Ga^{r t x} - i  q K_A (r;\om) \Ga^{x}  \ri) \ .
 \ee
Note that in~\eqref{eqn:foet} we have suppressed the boundary spinor indices which are restored here.

The effective vertex $\Lam$ can be evaluated explicitly. In particular
the integration over $r$ in~\eqref{effV} can be separated into $\Lam = \Lam_{IR} + \Lam_{UV}$, the contributions from the near horizon AdS$_2$ region ($\Lam_{IR}$) and from the rest of the spacetime ($\Lam_{UV}$). One finds that in the low temperature limit $\Lam_{UV} \sim O(1)$ while $\Lam_{IR} \sim T^{2 \nu_{k_F} +1}$ near the Fermi surface.

Since the effective vertex $\Lam$ can be approximated as a constant at the low temperature limit, the standard formula to express the
optical conductivity in terms of self-energy applies, \ie,~\eqref{eqn:foet} can be written as
 \be \label{cond}
 \sig (\om) = C'  \int {d \om' \ov 2 \pi}
 \,
  { f(\omega' + \om) - f(\omega') \over i \omega} \, {1 \ov {1 \ov v_F} \om + \Sig (\om' + \om) - \Sig^* (\om')}
 \ee
where $C'$ is a positive constant and the self-energy $\Sig$ is given in~\eqref{finiteTspinorG}.

\vspace{0.2in}   \centerline{\bf{Acknowledgements}} \vspace{0.2in}
We thank E.~Abrahams, P.~Coleman, S.~Hartnoll, G.~Horowitz, S.~Kachru, P.~Lee, S-S. Lee, J.~Maldacena, J.~Polchinski, K.~Rajagopal, S.~Sachdev, K.~Schalm,  E.~Silverstein, S.~Trivedi, C.~M.~Varma, J.~Zaanen and in particular T.~Senthil for discussions and encouragement.
Work supported in part by funds provided by the U.S. Department of Energy
(D.O.E.) under cooperative research agreement DE-FG0205ER41360 and the OJI program,
and in part by an Alfred P. Sloan fellowship.

\end{document}